\documentclass[a4paper,fleqn,usenatbib]{mnras}
\usepackage[T1]{fontenc}
\usepackage{ae,aecompl}

\usepackage{graphicx}
\usepackage{amsmath}
\usepackage{amssymb}
\usepackage{makecell}
\usepackage{bm}
\usepackage{xcolor}

\usepackage{aas_macros}

\definecolor{darkred}{rgb}{0.8, 0.0, 0.0}

\newcommand{\mathdash}{\,\text{--}\,}
\newcommand{\diff}[2]{{\frac{d{#1}}{d{#2}}}}
\newcommand{\pdiff}[2]{{\frac{\partial{#1}}{\partial{#2}}}}

\title[The role of atmospheric outflows in the migration]{The role of atmospheric outflows in the migration of hot Jupiters}

\author[E. P. Kurbatov and D. V. Bisikalo]{E. P. Kurbatov$^{1}$\thanks{E-mail: kurbatov@inasan.ru}
and D. V. Bisikalo$^{1}$
\\
$^{1}$Institute of Astronomy of the RAS, Moscow, Russia
}

\date{Accepted XXX. Received YYY; in original form ZZZ}

\pubyear{YYYY}

\begin{document}

\label{firstpage}
\pagerange{\pageref{firstpage}--\pageref{lastpage}}
\maketitle

\begin{abstract}
Many of observed hot Jupiters are subject to atmospheric outflows. Numerical simulations have shown that the matter escaping from the atmosphere can accumulate outside the orbit of the planet, forming a torus. In a few $10^8$ yr, the mass of the torus can become large enough to exert a significant gravitational effect on the planet. Accumulation of mass, in its own turn, is hindered by the activity of the star, which leads to the photoevaporation of the torus matter. We explore the role of these and other factors in the planet's migration in the epoch when the protoplanetary disk has already disappeared. Using HD~209458 system as an example, we show that the gravitational interaction with the torus leads to the possibility of migration of the planet to its observable position, starting from an orbit $\gtrsim 0.3$~AU.
\end{abstract}

\begin{keywords}
accretion, accretion disks -- planets and satellites: gaseous planets -- planet-disc interactions -- planets: migration
\end{keywords}



\section{Introduction}

A possible reason for the existence of gas giants in the orbits of $0.1$~AU and closer to their host stars, is migration of planets as a result of gravitational interaction with the protoplanet's disk. There are three types of migration, depending on the mass of the planet, disk density, and physical conditions in the gas \citep{Armitage2009apf..book.....A, Lubow2010arXiv1004.4137L}. The planets of several Earth mass are mainly subject to the Type~I migration, when the gravitational action of the planet on the disk excites tidal waves in the disk. The rate of exchange of orbital angular momentum is determined by the intensity of the planet--wave gravitational interaction \citep{Goldreich1980ApJ...241..425G, Artymowicz1993ApJ...419..155A}. For the planets with the mass of the order of Jupiter mass in the high-density gaseous disks, Type~II migration is usially assumed \citep{Lin1986ApJ...309..846L}, when the momentum of the force of the planet leads to the sweeping of the gas from the vicinity of the planet's orbit. As a result a gap is formed. Under the action of viscous forces in the gas, the gap becomes filled. The balance of both processes determines the rate of exchange of angular momentum. Other migration mechanisms are also possible, see, e.g., \citet{Nayakshin2012MNRAS.426...70N}. Intermediate mass planets may be subject to a mixed I+II scenario or participate in the Type~III migration (see \citet{Masset2003ApJ...588..494M} and references therein). If several planets are present in the disk simultaneously, the number of possible migration scenarios greatly increases.

Most of the classical T Tauri stars (young stars that have age of about $10^6$~yr) reveal observational manifestations of the gas--dust accretion disks \citep{Beckwith1990AJ.....99..924B}. Typical disks have a characteristic scale of the order of several hundred AU, their mass is several percent of the mass of the star. Observations \citep{Haisch2001ApJ...553L.153H}, as well as evolutionary models, show \citep{Bertout2007A&A...473L..21B, Galli2015A&A...580A..26G} that the lifetime of gaseous protoplanetary disks is, on average, several million years and does not exceed $10^7$~yr. The timescale of migration of a hot Jupiter from an orbit of $5$~AU to its typical orbit of $0.1$~AU can be as short as $10^5$~yr \citep{Armitage2009apf..book.....A} or even shorter \citep{Papaloizou2006RPPh...69..119P}. Of course, this time estimate depends on the density and viscosity of the disk, but it is much shorter than the lifeteime of the gas disk. In order to resolve the controversy between existing opinions on the formation of massive planets, as well as conflics with observations, various scenarios are proposed to slow down migration or even reverse it \citep{Lubow2010arXiv1004.4137L, Podlewska-Gaca2012MNRAS.421.1736P, Masset2003ApJ...588..494M}. In any case, it can be argued that the migration mechanisms are not completely  understood as yet.

The light curve of the system HD~209458 suggests that the atmosphere of the hot Jupiter HD~209458b is subject to the outflow \citep{Vidal-Madjar2003Natur.422..143V, Vidal-Madjar2008ApJ...676L..57V}. The outflow rate depends on the distance to the star, stellar activity, and chemical composition of the planet's atmosphere. Numerical 3D calculations performed by \citet{Shaikhislamov2020MNRAS.491.3435S} and \citet{Debrecht2020MNRAS.493.1292D} showed that, in the case of a relatively weak stellar wind, the matter escaping from the planet can accumulate outside planet's orbit, forming a torus. Hence, it is reasonable to assume that the mass accumulated over time can produce a noticeable gravitational effect on the orbit of the planet.

In our earlier paper \citep[][ henceforth, Paper~I]{Kurbatov2020ARep...64.1000K} we began to study the interaction of the accumulated matter and the planet. The model was based on the Pringle description of the accretion disk subject to a tidal torque. It was shown that the effect of tidal interaction between the planet and the torus becomes significant in the timescale of tens and hundreds of millions of years.

In this paper, we continue the research started in Paper~I. Therefore, the basic idea of the model remains the same: the matter leaving the atmosphere accumulates in an orbit close to the planet's orbit and interacts with the planet via tides, causing the migration of the orbit. The efficiency of the suggested migration mechanism is determined by the amount of accumulated matter and its distribution. In the present study we estimate various factors that can affect distribution of the matter in the torus and thus speed up or slow down migration and examine evolution of the orbit of a hot Jupiter as a result of the gravitational interaction of the planet and its ejecta.

\section{The factors affecting the matter distribution}
\label{sec:factors}

\subsection{The outflow of the planet's atmosphere}
\label{sec:outflow}

Many authors have estimated the rate of loss of the atmosphere of HD~209458b for the given flux of X-ray, EUV, and Ly$\alpha$ radiation. The process of atmosphere escape involves various physical factors, including thermal and ionization structure, thermal energy transfer, stellar wind \citep{Bisikalo2013ARep...57..715B}, radiation \citep{Cherenkov2018MNRAS.475..605C, Cherenkov2019ARep...63...94C}, and magnetic field \citep{Zhilkin2020ARep...64..159Z}; see also the recent review by \citet{Gronoff2020JGRA..12527639G}. Escape rate estimates vary considerably, from $10^{10}$~g\,s$^{-1}$ \citep{Vidal-Madjar2003Natur.422..143V} to $10^{12}$~g\,s$^{-1}$ \citep{Lammer2003ApJ...598L.121L}. \citet{Louden2017MNRAS.464.2396L} found, using an energy-limited escape model, that the mass-loss rate is $(3.8 \pm 0.2) \times 10^{10}$~g\,s$^{-1}$. The latter authors stated that the considerations of energy limit can provide the upper bound for the mass loss rate only. However, if chemical elements heavier than hydrogen are taken into account, alternative channels for the distribution of radiation energy become open. They lead to the change in the thermal structure of the planet's atmosphere and can change the estimate of the mass-loss rate. For instance, quite recently \citet{Lampon2020A&A...636A..13L} presented a 1D non-LTE model for a hydrogen-helium thermosphere of HD~209458b. The upper estimate for the escape rate was $10^{11}$~g\,s$^{-1}$. In the  earlier detailed calculations, \citet{GarciaMunoz2007P&SS...55.1426G} showed that the atmosphere of a hot Jupiter with an approximately solar composition (see for details the quoted paper) evaporates at a rate of up to $4.95 \times 10^{11}$~g\,s$^{-1}$.

In general, the escape rate can be written as $\dot{M}_\mathrm{p} \propto L_\mathrm{XUV}/a^2$, where $L_\mathrm{XUV}$ is stellar luminosity in the X- and EUV-ranges and $a$ is semi-major axis of the planet's orbit. This relation is good for $a > 0.015$~AU, while at closer distances the outflow is much more efficiently stimulated by the tidal action of the star \citep{GarciaMunoz2007P&SS...55.1426G}. X-ray and EUV luminosity of a post-ZAMS star decreases with the age approximately as $L_\mathrm{XUV} \propto t^{-1}$ \citep{Zahnle1982RvGSP..20..280Z}, so we can express the mass-loss rate of the planet in terms of some reference values:
\begin{equation}
  \label{eq:mass_loss_rate}
  \dot{M}_\mathrm{p}
  = \dot{M}_\mathrm{ref} \left( \frac{t}{t_\mathrm{ref}} \right)^{-1} \left( \frac{a}{a_\mathrm{ref}} \right)^{-2}  \;.
\end{equation}
As it is seen, we neglected the tidal force of the star.

Since there is no consensus in the literature on the escape rate, we should consider some limiting value for it: $\dot{M}_\mathrm{ref} = (4 \times 10^{10} \mathdash 5 \times 10^{11})$~g\,s$^{-1}$. We took the reference semi-major axis equal to $a_\mathrm{ref} = 0.047$~AU and the age of the star as $t_\mathrm{ref} = (3.5 \pm 1.4)\times10^9$~yr \citep{delBurgo2016MNRAS.463.1400D}.

\subsection{Photoevaporation of the torus}
\label{sec:photoevaporation}

\begin{figure*}
  \begin{center}
    \includegraphics{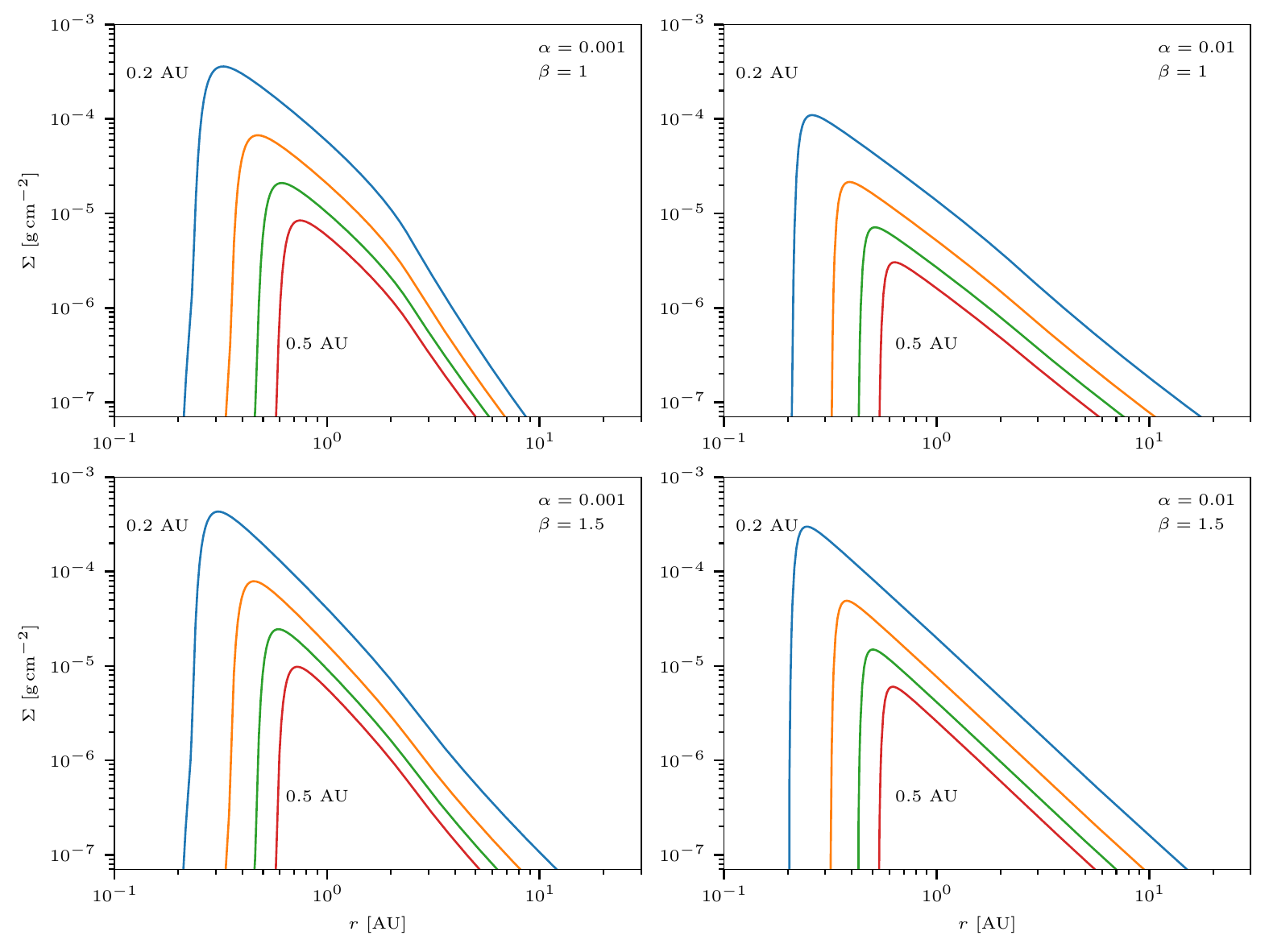}
  \end{center}
  \caption{Surface density profiles at $10^8$~yr from the start of the sumulations for different pairs of $\alpha$ and $\beta$ parameters and different initial orbits: $0.2$ (blue lines), $0.3$ (orange), $0.4$ (green), and $0.5$~AU (red).}
  \label{fig:grid-sigma-1e8}
\end{figure*}

The action of the ionizing radiation (EUV and X-ray) on the accretion disk and its subsequent photoevaporation is seen as the main cause for the disappearance of protoplanetary disks in the timescale of the order of $10^6$~yr \citep{Alexander2006MNRAS.369..216A}. Physical processes that control photoevaporation are, in principle, the same as for the atmospheric outflow, see above. The differences are in the centrifugal force coupled with the law of conservation of angular momentum \citep{Pereyra1997ApJ...477..368P}.

When a star is in the stage of a classical T Tauri one (CTT), its X-ray flux is comparable to or exceeds the EUV flux. However, the disk wind itself is optically thick to the EUV radiation, so it's presented only as a diffuse component \citep{Hollenbach1994ApJ...428..654H, Owen2010MNRAS.401.1415O}. At the same time, X-rays have a higher penetrating power and this leads to a much higher mass loss by the wind than in the case of EUV radiation alone \citep{Owen2012MNRAS.422.1880O}. This seems plausible also for the later stage of the evolution of a solar-type star, despite the fact that the X-ray luminosity declines faster over time than the EUV luminosity \citep{Tu2015A&A...577L...3T}.

Many authors in their analytical estimates have used the idea of a ``gravitational radius'' as an innermost radius in the disk where the photoevaporation can occur, see, e.g., \citet{Hollenbach1994ApJ...428..654H}. \citet{Liffman2003PASA...20..337L} showed that the correct expression for this quantity is
\begin{equation}
  \label{eq:gravitational-radius}
  r_\mathrm{g}
  = \frac{\gamma-1}{2\gamma}\,\frac{G M_\mathrm{s}}{c_\mathrm{s}^2}  \;,
\end{equation}
where $\gamma$ is the adiabatic index, $c_\mathrm{s}$ is the isothermal sound velocity. For a solar-mass star, monoatomic gas, and $c_\mathrm{s} = 10$~km\,s$^{-1}$ we have $r_\mathrm{g} = 1.8$~AU. Lifman's model was based on the Bernoulli equation for a compressible gas, applied to a geometrically thin disk. In the outer area of the disk, where $r > r_\mathrm{g}$, specific energy of the gas particles is positive, so the gas is unbound. In the inner area the situation is reverse. However, in the inner region, high enough above the disk surface, the energy criterion will  be met again and this will lead to the outflow from the inner region too. These ideas were confirmed by numerical calculations \citep{Ercolano2009ApJ...699.1639E, Owen2010MNRAS.401.1415O}: the star with $L_\mathrm{X} = 2\times10^{30}$~erg\,s$^{-1}$ is able to enable disk wind at the distances from $1$~AU and further \citep{Ercolano2009ApJ...699.1639E}.

In the papers cited above, their authors assumed that the protoplanet's disk is optically thick for ionizing radiation. In our formulation of the problem, gas torus is formed by the matter leaving the planet's atmosphere. According to our earlier calculations (Paper~I), the maximum of the surface density in the torus can be as high as $\Sigma_\mathrm{max} \approx 0.04$~g\,cm$^{-2}$. This amount of the matter has a large optical thickness in EUV, $\gtrsim 10^5$, but rather moderate thickness in X-ray. Indeed, assume that $1$~keV photons dominate in the X-ray flux \citep{Ercolano2009ApJ...699.1639E}, then the cross-section is about $\sigma_\mathrm{X} = 2\times 10^{-22}$~cm$^{-2}$ \citep{Ride1977A&A....61..339R}. As a result, the maximum optical thickness will be $\sigma_\mathrm{X} \Sigma_\mathrm{max}/m_\mathrm{H} \approx 5$. In the outer parts of the torus, where the density is small and/or at earlier epochs, when not much mass has already accumulated, the torus may be optically thin.

Taking into account everything written above as a basis for the photoevaporation model, we take the approximation of \citet{Owen2012MNRAS.422.1880O} which they suggested for the accretion disks with inner holes. After proper normalization, the basic exspression for the surface mass loss rate may be written as
\begin{multline}
  \label{eq:photoevaporation-rate-basic}
  \dot{\Sigma}_\mathrm{pe}^{(0)}
  = \left( 2\times10^{-10} \text{~g\,cm$^{-2}$\,s$^{-1}$} \right)
    \left( \frac{M_\mathrm{s}}{M_\odot} \right)^{-1.148}  \\
    \times \left( \frac{L_\mathrm{X}}{10^{30} \text{~erg\,s$^{-1}$}} \right)^{1.14}
    \left( \frac{r}{1 \text{~AU}} \right)^{-1} F(y)  \;,
\end{multline}
where $F(y) = a b \exp(b y) + c d \exp(d y) + e f \exp(f y)$, and the symbols $a$ to $f$ are fitting parameters, see Appendix B2 in \citet{Owen2012MNRAS.422.1880O}. The value of $y$ is the measure of the distance from the inner hole edge $r_0$:
\begin{equation}
  y = 0.95 \left( \frac{M_\mathrm{s}}{M_\odot} \right)^{-1} \frac{r - r_0}{1 \text{~AU}}  \;,\qquad
  r \geq r_0  \;.
\end{equation}

For the timescales $\sim 10^8$~yr and longer, the time dependence of the X-ray luminosity becomes important. According to \citet{Tu2015A&A...577L...3T}, X-ray luminosity of the Sun (G2V star) depends on time as $L_\mathrm{X} \propto t^{-1.42}$.  Current X-ray luminosity of HD~209458 (G0V star) is $L_\mathrm{X,ref} = 10^{28.08 \pm 0.07}$ \citep{Louden2017MNRAS.464.2396L}. As a result, we obtain
\begin{equation}
  \label{eq:x-ray-luminosity}
  L_\mathrm{X}
  = L_\mathrm{X,ref} \left( \frac{t}{t_\mathrm{ref}} \right)^{-1.42}  \;.
\end{equation}

Model \eqref{eq:photoevaporation-rate-basic} is formally correct for any $r_0$, although the authors tested it for $r_0 \geq 5.7$~AU and solar-mass star \citep{Owen2011MNRAS.412...13O}, as well as for $r_0 = 0.7$~AU and $0.1\,M_\odot$ star \citep{Owen2012MNRAS.422.1880O}. We will apply energy considerations after \citet{Liffman2003PASA...20..337L} and take $r_\mathrm{g}$ from Eq.~\eqref{eq:gravitational-radius} as an innermost radius for photoevaporation. We also take into account, in a very simple manner, that the torus can be optically thin to X-ray radiation. The final expression for the photoevaporation rate is
\begin{equation}
  \label{eq:photoevaporation-rate}
  \dot{\Sigma}_\mathrm{pe}
  = \dot{\Sigma}_\mathrm{pe}^{(0)} \left[ 1 - e^{-\kappa_\mathrm{X} \Sigma} \right]
\theta(r > r_\mathrm{g})  \;,
\end{equation}
where $\kappa_\mathrm{X} = \sigma_\mathrm{X}/m_\mathrm{H} = 119.5$~cm$^2$\,g$^{-1}$; $\theta(\cdot)$ is the Heaviside step function, which is unity if the condition is satisfied and zero otherwise.

\subsection{Stellar wind}
\label{sec:stellar-wind}

\begin{figure}
  \begin{center}
    \includegraphics{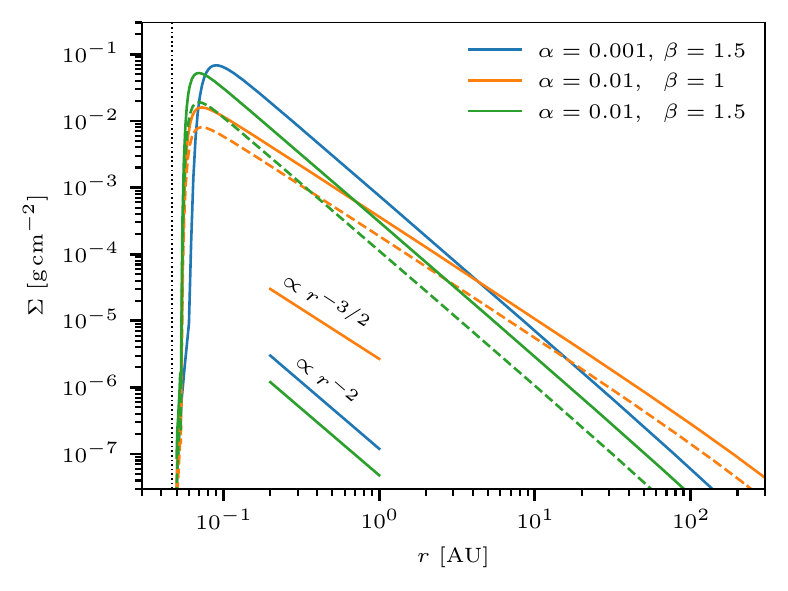}
  \end{center}
  \caption{Surface density profiles at the end of simulations, when the semi-major axis of the orbit becomes $a_\mathrm{ref} = 0.047$~AU. {\em Solid lines} show models, in which $a_\mathrm{ini} = 0.2$~AU. {\em Dashed lines} show the models, in which $a_\mathrm{ini} = 0.3$~AU. The dotted vertical line shows position of the planet. }
  \label{fig:sigma-fin}
\end{figure}

Stellar wind particles can interact with the gas torus and to be a source of additional radial force acting on the gas. This becomes important if the dynamic pressure of the wind exceeds the gas pressure in the torus:
\begin{equation}
  \label{eq:wind-dominance-condition}
  m_\mathrm{p} N_\mathrm{w} v_\mathrm{w}^2 \gtrsim \rho c_\mathrm{s}^2  \;,
\end{equation}
where $N_\mathrm{w}$ and $v_\mathrm{w}$ are number density and velocity of the wind particles, correspondingly; $\rho$ is the density at the torus' inner edge; $c_\mathrm{s}$ is the sound velocity. It is not easy to take the wind into account in the Pringle's model. Instead, let us find out the role of the wind by some simple estimates.

\citet{Withbroe1988ApJ...325..442W} suggested a one-fluid model of the solar wind, both for the quiet Sun and for its active regions. According to this model, dynamic pressure of the wind of the quiet Sun is of the order of $10^{19}$~cm$^{-1}$\,s$^{-2}$, weakly depending on the distance for $r < 10~R_\odot = 0.047$~AU. Over the same distance interval, dynamic pressure of the wind from coronal holes quickly drops from $10^{23}$ to $10^{14}$~cm$^{-1}$\,s$^{-2}$. According to \citet{Parker1958ApJ...128..664P} model for coronal temperature $2\times10^6$~K, at the distances from $10~R_\odot$ to $1$~AU wind pressure of the quiet Sun decreases from $10^{19}$ to $4\times 10^{16}$~cm$^{-1}$\,s$^{-2}$.

On the other hand, in our Paper~I we found that, when the planet is at the orbit of $0.047$~AU, the surface density at the torus inner edge is typically $\gtrsim 0.003$~g\,cm$^{-2}$. Given the sound speed $10$~km\,s$^{-1}$ and vertical equillibrium condition at a distance $10~R_\odot$, the number density of the torus matter is of the order of $10^{10}$~cm$^{-3}$. In this case the r.h.s. of the Eq.~\eqref{eq:wind-dominance-condition} is $10^{22}$~cm$^{-1}$\,s$^{-2}$, much greater than the wind dynamic pressure.

It should be said that as a result of photoevaporation, the number density in the torus can decrease and the condition \eqref{eq:wind-dominance-condition} will weaken. However, later we will see that it was quite justified to neglect the effect of the stellar wind.

\subsection{Viscosity}

\begin{figure*}
  \begin{center}
    \includegraphics{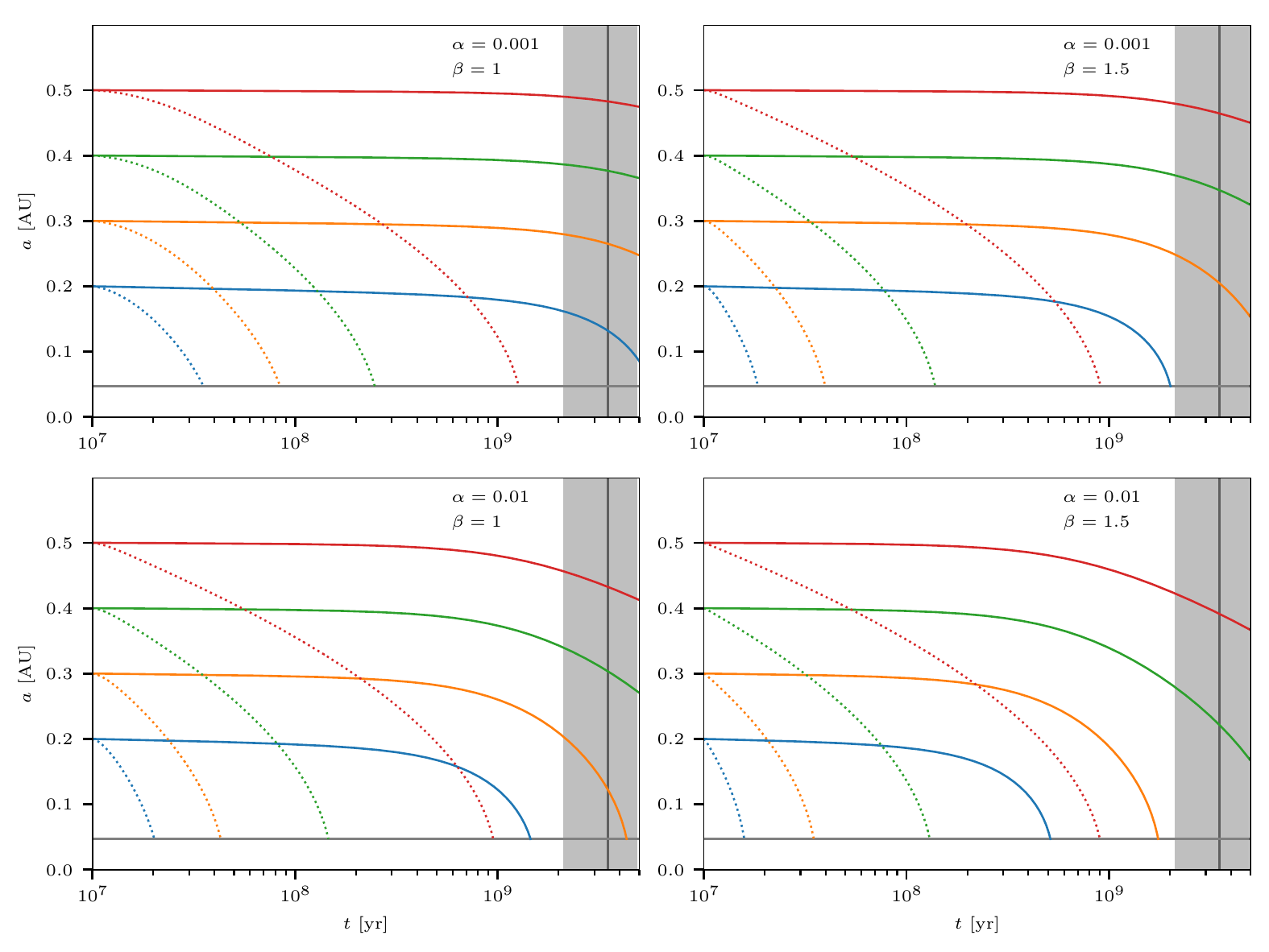}
  \end{center}
  \caption{Semi-major axis of the planet's orbit as a function of stellar age. {\em Dotted lines} show results of simulations without photoevaporation (i.e., the model from the Paper~I). {\em Solid lines} show results of simulations according to the model from this paper. Positions of the gray vertical line and the bar correspond to the estimated age of the star $(3.5\pm 1.4)\times10^9$~yr \citep{delBurgo2016MNRAS.463.1400D}. Gray horizontal line indicates the final orbit of the planet $a_\mathrm{ref} = 0.047$~AU.}
  \label{fig:grid-a}
\end{figure*}

In the non-magnetic astrophysical disks, turbulent viscosity is the most efficient mechanism for the transfer of the angular momentum. Then the hydrodynamic instabilities are the most likely source of the turbulence in such systems. It is often believed that an instability in a non-magnetic gaseous disk can develop only in an entropy-driven way \citep{Armitage2015arXiv150906382A}. For instance, if the radiative cooling is not efficient enough for the given heating rate, the disk becomes convectively unstable. Convection, in its own turn, is a source of the turbulence \citep{Canuto1997ApJ...482..827C}. If the disk is efficiently cooled radiatively, but has a vertical shear of azimuthal velocity (which is inevitable in the case of the radial temperature gradient), the vertical shear instability develops. Numerical simulations of the protoplanetary disks have shown, however, that the cooling agent should be as effective as dust. If there is little amount of dust, then this type of instability is unlikely to develop \citep{Lin2015ApJ...811...17L}. Also note that long before the stage that is of interest to us, a significant fraction of the dust has already condensed and does not longer contribute to the heat balance of the gas. However, in Paper~I we have shown than the torus is optically thin in continuum. Indeed, surface density in the torus has upper limit of $0.04$~g\,cm$^{-2}$. Assuming full ionization (which is certainly not true), the optical thickness of the torus with respect to the electron scattering ($\kappa^\mathrm{es} \approx 0.4$~cm$^2$\,g$^{-1}$) does not exceed $0.016$. The free-free opacity, $\kappa^\mathrm{ff} \approx 0.11\,N/T^{7/2}$~cm$^2$\,g$^{-1}$, is several orders of magnitude lower than $\kappa^\mathrm{es}$, for $N = 10^{11}$~cm$^{-3}$ and $T = 10^4$~K. Therefore, one can conclude that the torus has a very short cooling time, about $(c \kappa^\mathrm{es} m_\mathrm{H} N)^{-1} < 6\times10^{-3}$~day. This favours the vertical shear instability.

Another possible way to induce the turbulence is an instability owing to the external disturbances. Such disturbances are the tidal waves, excited by the planet, as well as hydrodynamic perturbations that arise where the stream of the outflowing atmosphere is in the contact with the inner edge of the torus. An obvious mechanism for the growth of disturbances in this case is the Kelvin-Helmholtz instability. Alternative ways for an instability, which can arise even in a purely isothermal matter distribution, are: (i) overlap of the phases of perturbations or (ii) overturn of the profile of the perturbation. The first mechanism requires a rather strong and extended background disturbance, more precisely, the variance of the background velocity over the wavelength scale must exceed the sound velocity \citep{Kurbatov2014PhyU...57..851K, Kurbatov2018ARep...62..781K}. This is clearly not the case for the tidal waves, because they are subsonic \citep{Goldreich1979ApJ...233..857G}. The second mechanism based on a nonlinear effect that occurs for almost any finite-amplitude perturbation is the wave profile overturn and  transformation of the wave into a shock \citep{Landau1959flme.book.....L}, which, in turn, can be unstable. This scenario is quite natural in circumstellar disks of cataclysmic binary stars where the inner edge of the disk is subject to strong disturbances, both gravitational and hydrodynamic \citep{Kurbatov2017ARep...61..475K, Kurbatov2017ARep...61.1031K}. As it is seen, it can be argued that the turbulence in the gas torus is quite probable.

Finally, one can offer an evolutionary argument for turbulence in protoplanet's disks. Formation time of a planet of Jupiter type depends on the gas viscosity, as it determines the amount of matter in the disk available for accretion onto the planet (in the core-collapse model, see, e.g., \citet{Bodenheimer2013ApJ...770..120B} and references therein). Tidal action of the planet leads to the sweeping of the gas from the vicinity of its orbit. Too low viscosity will lead to too early opening of the gap in the disk and termination of the growth of the planet's mass.

Turbulent viscosity can be parametrized by the spatial correlation scale $\ell$ and the correlation time $\tau$ as $\nu = \ell^2/\tau$. In $\alpha$-notation of \citet{Shakura1973A&A....24..337S}, viscosity has the form $\nu = \alpha H^2 \Omega$. The dimensionless coefficient $\alpha$ can be expressed as
\begin{equation}
  \label{eq:alpha-general}
  \alpha
  = \left( \frac{\ell}{H} \right)^{\!2} \frac{1}{\Omega \tau}  \;.
\end{equation}
The characteristic time of the azimuthal flow in the disk, as well as of the vertical sound oscillations, is of the order of $\Omega^{-1}$. This is also the timescale for the propagation of the turbulence in the azimuthal direction, i.e., its correlation time. Hence, we may set $\Omega \tau = 1$ in Eq.~\eqref{eq:alpha-general}. On the other hand, the vertical disk scale is often considered as the turbulence correlation scale, as if it were free turbulence in a planar jet \citep{Rodi1970W&S.....3...85R}. In this case, $\alpha \equiv \operatorname{const}$ and $\nu \propto H^2 \Omega \propto T/\Omega$, where $T$ is the gas temperature. If it suddenly turns out that the reasoning about the time and the scale of the correlation is not correct, we will take a general power-law expression for the viscosity:
\begin{equation}
  \label{eq:viscosity-coeff}
  \nu
  = \alpha H_0^2 \Omega_0 \left( \frac{r}{r_0} \right)^\beta  \;,
\end{equation}
where $H_0$ and $\Omega_0$ are semi-thickness and angular frequency at a certain reference point $r_0$.

Note that according to the recent observations of protoplanet's disks, the upper estimates of the viscosity parameter are $\alpha \lesssim 0.003 \mathdash 0.007$ \citep{Flaherty2017ApJ...843..150F, Flaherty2018ApJ...856..117F}.

Below, in the model, we try various combinations of $\alpha = 0.001$, $\alpha = 0.01$, $\beta = 1$, and $\beta = 3/2$. The case $\beta = 3/2$ correstponds to the isothermal distribution of the gas along the radius in a Keplerian disk, and the case $\beta = 1$ corresponds to the case when the temperature decreases as $T \propto r^{-1/2}$.

\subsection{Tidal interaction}

Tidal interaction between the planet and the disk leads to the redistribution of angular momentum outward \citep{Goldreich1980ApJ...241..425G}. Interaction is the strongest in the immediate vicinity of the planet's orbit, so the gas distribution in this area determines the intensity of the exchange of angular momentum. The action of tides causes the gas to be swept out and a gap is formed. The viscous force, in its own turn, tends to fill the gap.

If we are interested in the planet-disk interaction in the timescale much larger than the orbital period of the planet, then it will be sufficient to use an approximate expression for tidal torque \citep{Goldreich1980ApJ...241..425G, Papaloizou2006RPPh...69..119P}:
\begin{equation}
  \label{eq:tidal-force-density-general}
  \tau
  = \frac{C_0}{\pi}\,\frac{G^2 M_\mathrm{p}^2 a}{(r - a)^2}
    \,\frac{\Omega_\mathrm{p} - \Omega}{(a \Omega_\mathrm{p} - r \Omega)^3}  \;,
\end{equation}
where $C_0 \approx 2.82$. Here, the planet has a circular orbit of the semi-major axis $a$ and Keplerian angular velocity $\Omega_\mathrm{p}$. Angular velocity of the gas is $\Omega = \Omega(r)$.

The positive sign of the r.h.s. in the Eq.~\eqref{eq:tidal-force-density-general} implies that $r > a$, i.e., the planet loses its angular momentum to the disk. This occurs at the rate
\begin{equation}
  \label{eq:angular-momentum-conservation}
  \diff{}{t} (M_\mathrm{p} a^2 \Omega_\mathrm{p})
  = - 2\pi \int_a^\infty dr\,r \Sigma \tau  \;.
\end{equation}

\section{The model of the orbit migration}

Let's collect everything written above into one model. We take as a basis the Pringle model, where the mass transfer is controlled by the transfer of angular momentum \citep{Pringle1981ARA&A..19..137P}.
In comparison to the original model, the tidal torque is added to the angular momentum equation and the photoevaporation is added to both the angular momentum and the continuity equation:
\begin{gather}
  \label{eq:mass-transfer-general}
  \pdiff{\Sigma}{t} + \frac{1}{r}\,\pdiff{(r F)}{r}
  = - \dot{\Sigma}_\mathrm{pe}  \;,  \\
  \label{eq:angular-momentum-transfer-general}
  \pdiff{}{t} (\Sigma r^2 \Omega) + \frac{1}{r}\,\pdiff{(r F r^2 \Omega)}{r}
  = \frac{1}{r}\,\pdiff{(r^2 W)}{r}
  - \dot{\Sigma}_\mathrm{pe}\,r^2 \Omega
  + \Sigma \tau  \;.
\end{gather}
Here $F$ is the radial mass flux density and $W$ is `$r\phi$' component of the viscosity stress tensor,
\begin{equation}
  \label{eq:reynolds-tensor-general}
  W
  = \nu \Sigma r\,\pdiff{\Omega}{r}  \;,
\end{equation}
where the turbulent viscosity coefficient is defined by Eq~\eqref{eq:viscosity-coeff}.

Let the angular velocity profile $\Omega(r)$ to be time-independent. Then the flux can be expressed by means of Eq.~\eqref{eq:angular-momentum-transfer-general}, using Eq.~\eqref{eq:mass-transfer-general}:
\begin{equation}
  \label{eq:mass-flux-general}
  F
  = \left[ \pdiff{(r^2 \Omega)}{r} \right]^{-1}
    \left\{
      \frac{1}{r}\,\pdiff{(r^2 W)}{r}
      + \Sigma \tau
    \right\}  \;.
\end{equation}
Note that the photoevaporation term on the r.h.s. of~\eqref{eq:angular-momentum-transfer-general} is canceled in \eqref{eq:mass-flux-general} because the local specific angular momentum of the wind is the same as one in the torus.

Suppose, there is a Keplerian distribution of the gas angular velocity, $\Omega = (GM/r^3)^{1/2}$, then the mass flux is
\begin{equation}
  \label{eq:mass-flux}
  F
  = - \frac{3}{r^{1/2}}\,\pdiff{}{r}\!\left( r^{1/2} \nu \Sigma \right)
    + 2 \left( \frac{r}{G M_\mathrm{s}} \right)^{1/2}\Sigma \tau  \;,
\end{equation}
and
\begin{equation}
  \tau
  = \frac{C_0}{\pi}\,\frac{M_\mathrm{p}}{M_\mathrm{s}}\,\frac{G M_\mathrm{p} a}{(r - a)^2}
    \,\frac{r^{3/2} - a^{3/2}}{(r^{1/2} - a^{1/2})^3}  \;.
\end{equation}
Tidal torque exerted by the planet on the disk (the second term in the r.h.s. of the Eq.~\eqref{eq:mass-flux}) leads to an additional non-diffusive and non-negative flux, which, however, rapidly decreases, as $\Sigma/r^{3/2}$ for $r \gg a$.

Continuity equation \eqref{eq:mass-transfer-general} with the flux \eqref{eq:mass-flux} needs a pair of the boundary conditions. It can be assumed that the outflow of the atmosphere of the planet begins at the distance of the Hill radius from it. Given that the characteristic time of angular momentum exchange is much longer than the orbital period of the planet, we can place the mass source for the torus in an orbit with radius $r_0 = a + r_\mathrm{Hill}$. The mass flux is then
\begin{equation}
  \label{eq:mass-flux-boundary}
  F_0
  = F(r_0)
  = \frac{\dot{M}_\mathrm{p}}{2\pi r_0}  \;,
\end{equation}
where $\dot{M}_\mathrm{p}$ is defined in Eq.~\eqref{eq:mass_loss_rate}. The second boundary condition is zero surface density at infinity.

Migration velocity of the orbit of the planet is determined by Eq.~\eqref{eq:angular-momentum-conservation}. Since the change of the planet mass can be neglected, migration rate is
\begin{equation}
  \diff{a}{t}
  = - \frac{4\pi}{M_\mathrm{p}} \left( \frac{a}{G M_\mathrm{s}} \right)^{1/2} \int_{r_0}^\infty dr\,r \Sigma \tau  \;.
\end{equation}
Of course, formal infinity at the upper limit of integration denotes, in fact, some rather distant point.

The last expression, together with initial conditions for the surface density and the semi-major axis, closes the system of equations \eqref{eq:mass-transfer-general} and \eqref{eq:mass-flux}--\eqref{eq:mass-flux-boundary}. The domain area of the system is $r_0 \leqslant r < \infty$. Since the semi-major axis $a$ changes over time and, hence, the inner boundary of the domain $r_0$, it is convenient to apply the Lagrange moving grid for the numerical solution in this domain. In order to do this, we introduce the characteristic scales:
\begin{gather}
  t_0 = \frac{1}{\Omega_0}  \;,\qquad
  \Sigma_0 = \frac{\dot{M}_\mathrm{p}}{2\pi r_0^2 \Omega_0}  \;,  \\
  \nu_0 = r_0^2 \Omega_0  \;,\quad
  \Omega_0 = \Omega(r_0)
\end{gather}
and give the equations dimensionless form: $s = t/t_0$, $x = r/r_0$, $\sigma = \Sigma/\Sigma_0$, $f = F/F_0$, $n = \nu/\nu_0$, $\xi = a/r_0$, $h = H_0/r_0$,
\begin{gather}
  \label{eq:mass-transfer-unitless}
  \pdiff{\sigma}{s} + \frac{1}{x}\,\pdiff{(x f)}{x}
  = - \frac{t_0}{\Sigma_0} \left( \dot{\Sigma}_\mathrm{pe} + \diff{\Sigma_0}{t}\,\sigma \right)  \;,  \\
  \label{eq:mass-flux-pringle-unitless}
  f = - \frac{3}{x^{1/2}}\,\pdiff{}{x}\!\left( x^{1/2} n \sigma \right)
    + \xi^{1/2} x^{1/2} \omega \sigma  \;,  \\
  \label{eq:viscosity-coeff-unitless}
  n = \alpha h^2 x^\beta  \;,  \\
  \label{eq:specific-torque-unitless}
  \omega
  = \frac{2C_0}{\pi} \left( \frac{M_\mathrm{p}}{M_\mathrm{s}} \right)^2
      \frac{\xi^{1/2}}{(x - \xi)^2}
      \,\frac{x^{3/2} - \xi^{3/2}}{(x^{1/2} - \xi^{1/2})^3}  \;,  \\
  \diff{a}{s}
  = - a\,\frac{t_0 \dot{M}_\mathrm{p}}{M_\mathrm{p}} \int_1^\infty dx\,x \omega \sigma  \;.
\end{gather}
The new domain is now $1 \leqslant x < \infty$ and the boundary conditions are
\begin{equation}
  \label{eq:boundary-conditions-unitless}
  f\bigr|_{x=1} \equiv 1  \;,\qquad
  \sigma\bigr|_{x\to\infty} \equiv 0  \;.
\end{equation}
After the dimensionless soulution $\sigma(s, x)$ has been found, the physical solution can be expressed as $\Sigma(t, r) = \Sigma_0\,\sigma(t/t_0, r/r_0)$.

As it is seen, in Eq.~\eqref{eq:mass-transfer-unitless}, a new term has appeared on the r.h.s. This is due to the fact that the characteristic density scale $\Sigma_0$ depends on the time via $\dot{M}_\mathrm{p}$ and $r_0$. This term can be written as follows:
\begin{equation}
  \label{eq:reference-frame-drift}
  \frac{t_0}{\Sigma_0}\,\diff{\Sigma_0}{t}
  = t_0\,\pdiff{\ln \dot{M}_\mathrm{p}}{t}
    + \left[ \pdiff{\ln \dot{M}_\mathrm{p}}{a} - \diff{\ln (r_0^2 \Omega_0)}{a} \right] \diff{a}{s}  \;.
\end{equation}
Strictly speaking, two more terms should appear there:
\begin{equation}
  \diff{\ln t_0}{\ln s}\,\pdiff{\sigma}{s}  \qquad
  \text{and}  \qquad
  \frac{x}{s}\,\diff{\ln r_0}{\ln s}\,\pdiff{\sigma}{x}  \;.
\end{equation}
It can be shown, however, that in the timescale of the planet's migration (by $\sim 10$ orders of magnitude larger than the orbital period), these terms can be neglected.

This model can be slightly simplified due to the fact that in the photoevaporation area the gas density is sufficiently low for the torus to be optically thin. In such a case the exponential term in Eq.~\eqref{eq:photoevaporation-rate} can be linearized as $1 - e^{-\kappa_\mathrm{X} \Sigma} \approx \kappa_\mathrm{X} \Sigma$. At this point the model becomes linear in $\Sigma$. Hence, to solve it numerically, a standard discretization scheme can be applied to the Laplace-like differential operator at the l.h.s. of \eqref{eq:mass-transfer-unitless}, so the equation will be redefined to a spatial grid $\{x_i\}_{i=1}^N$. After this the boundary conditions \eqref{eq:boundary-conditions-unitless} should be redefined also. The first condition will remain unchanged, but the second one, $\sigma\bigr|_{x\to\infty} \equiv 0$, can be simulated as a free flow condition:
\begin{equation}
  \label{eq:boundary-condition-free-flow}
  x_{N-1} f_{N-1} = x_N f_N  \;,
\end{equation}
where $f_N$ is the mass flux at the interface $x_N$ of the rightmost cell.

\section{Application to HD~209458}

In the previous section we presented the model of planet -- torus tidal interaction taking into account the photoevaporation and planet's orbit migration. The basic system of equations of the model is \eqref{eq:mass-transfer-unitless}\mathdash\eqref{eq:reference-frame-drift}, and the additional equations are \eqref{eq:mass_loss_rate}, and \eqref{eq:gravitational-radius}\mathdash\eqref{eq:photoevaporation-rate}. In the model we use the parameters of HD~209458 from \citet{delBurgo2016MNRAS.463.1400D}. All the model parameters are summarized in Table~\ref{tbl:parameters}.
\begin{table}
  \centering
  \begin{tabular}{lrc}
    \hline
    \hline  \\[-3mm]
    Stellar mass & $M_\mathrm{s}$ & $1.148~M_\odot$  \\
    \makecell[lb]{Current X-ray luminosity \\
      \hspace{1cm} of the star} & $L_\mathrm{X,ref}$ & $10^{27.08}$ erg\,s$^{-1}$  \\
    Stellar age & $t_\mathrm{ref}$ & $4.9\times10^9$ yr  \\
    Planet mass & $M_\mathrm{p}$ & $0.74~M_\mathrm{Jup}$  \\
    \makecell[lb]{Current semi-major axis of the   \\
      \hspace{1cm} planet orbit } & $a_\mathrm{ref}$ & $0.047$ AU  \\
    \makecell[lb]{Current planet's atmosphere  \\
      \hspace{1cm} outflow rate } & $\dot{M}_\mathrm{ref}$ & $5\times10^{11}$ g\,s$^{-1}$  \\
    Gas sound velocity & $c_\mathrm{s}$ & $9.08$ km\,s$^{-1}$  \\
    Gas adiabatic exponent & $\gamma$ & $5/3$  \\
    \makecell[lb]{Stellar age at the beginning \\
      \hspace{1cm} of simulation} & $t_\mathrm{ini}$ & $10^7$ yr  \\
    \\[-3mm]
    \hline  \\[-3mm]
    \makecell[lb]{Initial semi-major axis of the \\
      \hspace{1cm} planet orbit} & $a_\mathrm{ini}$ & $0.2 \mathdash 0.5$ AU  \\
    Gas viscosity parameter & $\alpha$ & $0.001$, $0.01$  \\
    Viscosity profile exponent & $\beta$ & $1$, $1.5$  \\
    \\[-3mm]
    \hline
  \end{tabular}
  \caption{Upper part of the table ---  constant parameters of the model.  Bottom part --- varying parameters of the grid of models. See references in the text.}
  \label{tbl:parameters}
\end{table}

\begin{figure}
  \begin{center}
    \includegraphics{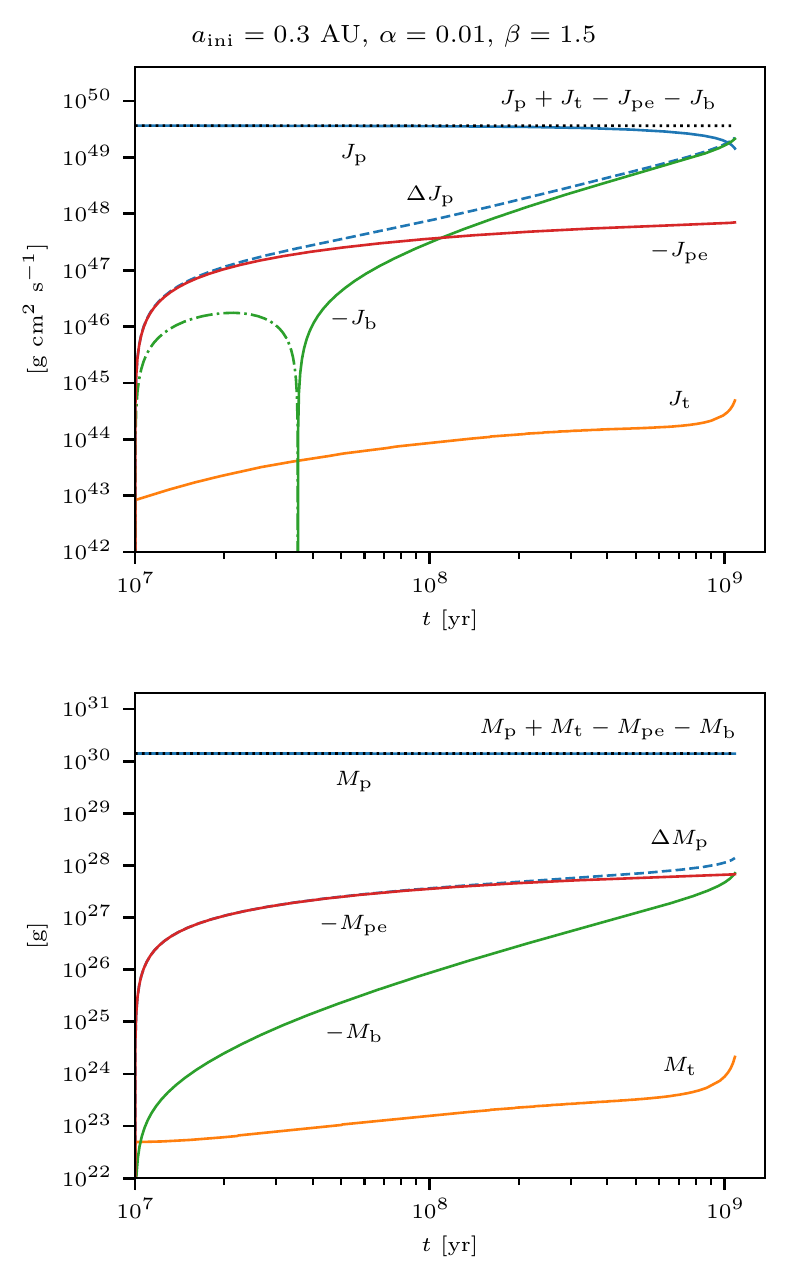}
  \end{center}
  \caption{Conservation of angular momentum (top panel) and mass (bottom panel) during simulations. Top panel: {\em blue line} is the planet's orbital momentum, $J_\mathrm{p}$; {\em dashed blue line} is the decrement of the angular momentum of the planet; {\em orange} is the angular momentum accumulated in the torus, $J_\mathrm{t}$; {\em red} is the cumulative momentum loss by the photoevaporation in the torus, $-J_\mathrm{pe}$; {\em Green} is the momentum transferred away through the boundaries, $-J_\mathrm{b}$ (dash-dots are for negative values); the {\em dotted line} is the sum of all the components. In the bottom panel the color lines denote the same contributions, see Eq.~ \eqref{eq:total-angular-momentum-conservation} and \eqref{eq:total-mass-conservation}.}
  \label{fig:conservation}
\end{figure}

\begin{figure}
  \begin{center}
    \includegraphics{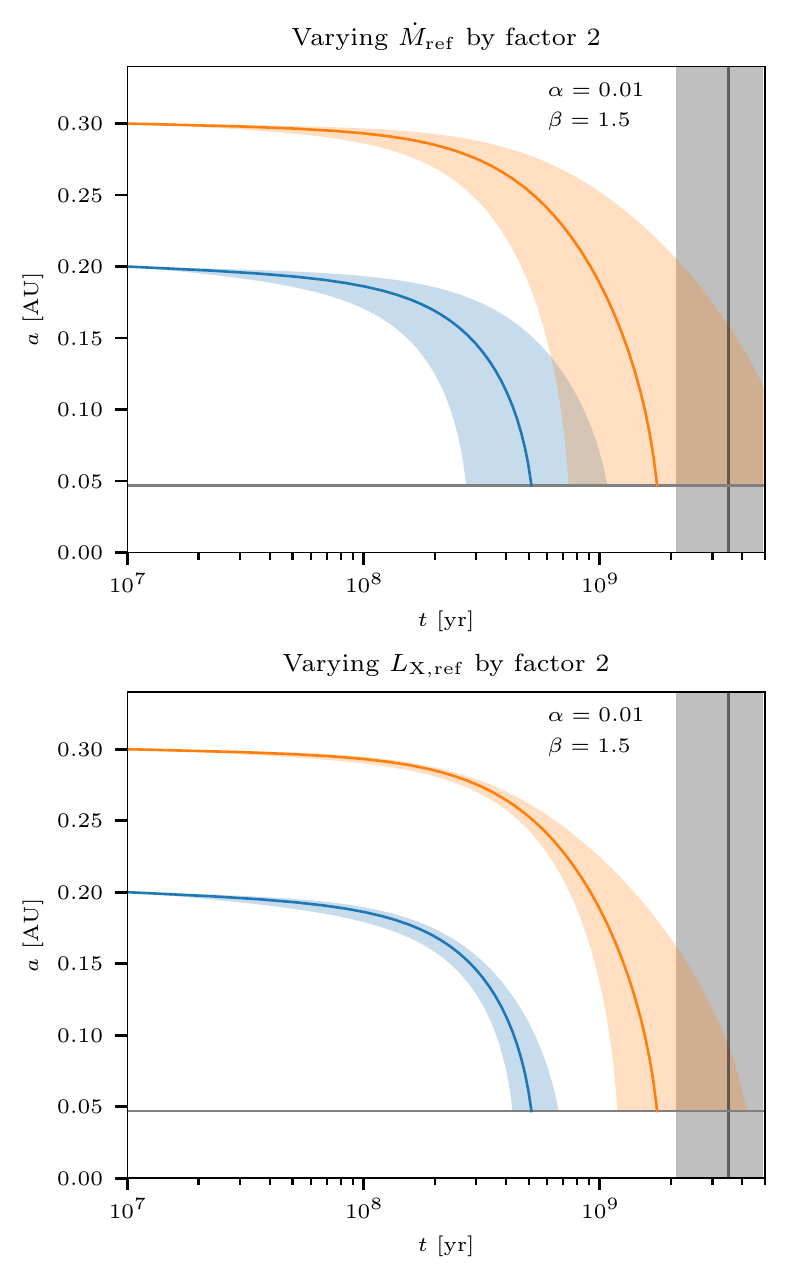}
  \end{center}
  \caption{Dependence of the evolution of the planet's orbit on the variation of atmosphere outflow rate (top panel) or stellar X-ray luminosity (bottom panel). Only case of the strong turbulence is shown. Coloured lines denote the same as in Fig.~\ref{fig:grid-a}. Blue and orange shaded areas show the boundaries of the evolutionary trajectory changes with variation of the parameters.}
  \label{fig:a-var}
\end{figure}

\begin{figure*}
  \begin{center}
    \includegraphics{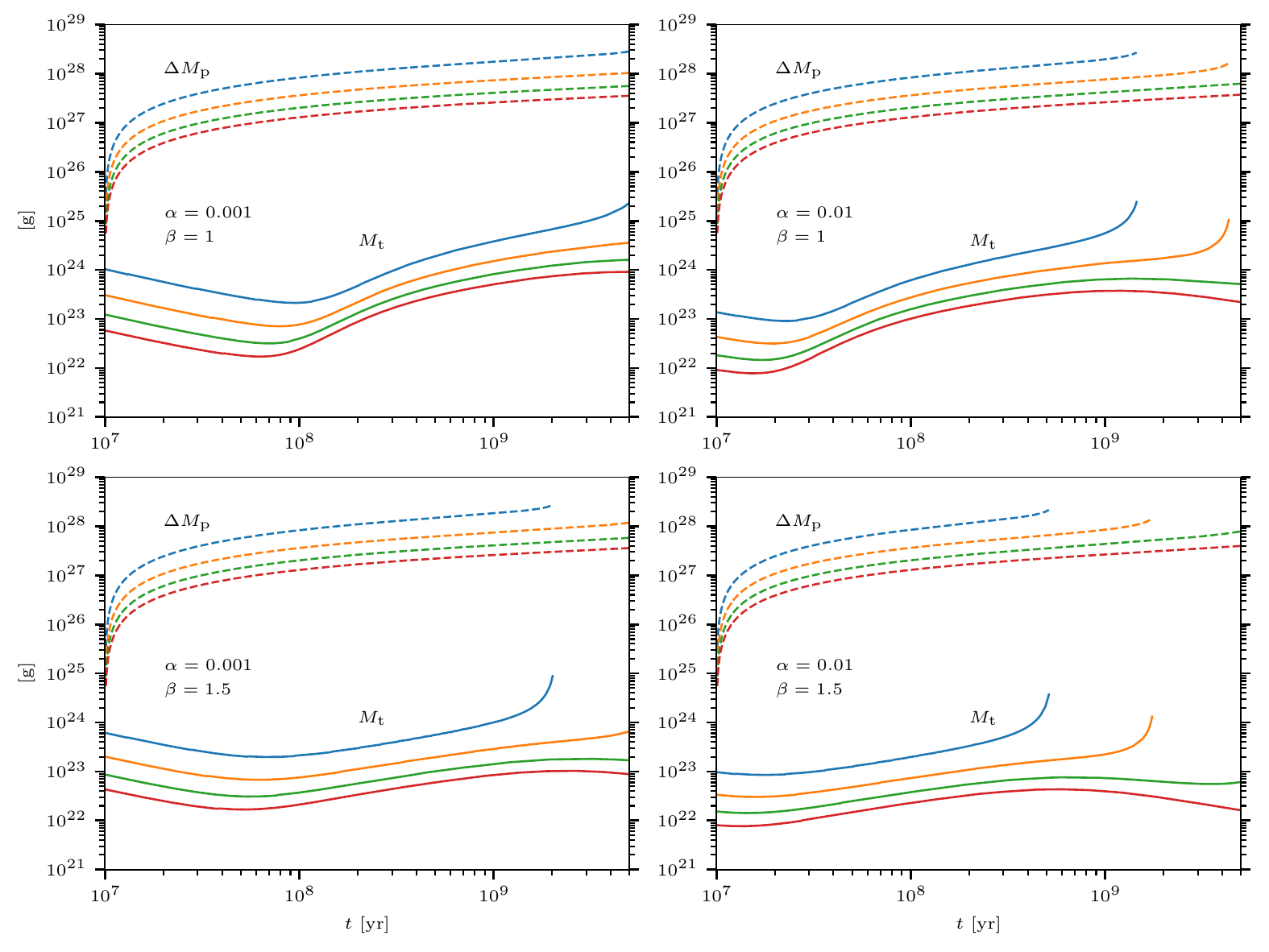}
  \end{center}
  \caption{The mass escaping the atmosphere of the planet (the lines at the top of each plot) and the torus mass (lines at the bottom). Colours of the lines have the same meaning as in Fig.~\ref{fig:grid-a}.}
  \label{fig:grid-mass}
\end{figure*}

Differential equations \eqref{eq:mass-transfer-unitless} and \eqref{eq:mass-flux-pringle-unitless} were solved numerically on a logarithmic grid of $1000$ cells for $1 \leq x \leq (10^4 \text{~AU})/r_0$. Time integration was carried out according to an implicit scheme%
\footnote{The numerical code as available at \url{https://github.com/evgenykurbatov/kb21-hotjup-migration-adv}} (see Paper~I).

We carried out a series of simulations, varying initial orbit of the planet, as well as the gas viscosity parameter and viscosity exponent, both in the order to cover the uncertainty of our understanding of the generation and propagation of the turbulence. The values of the viscosity parameter $\alpha$ that we applied correspond to the observed estimates of the turbulence intensity in protoplanetary disks. The simulations started at the age of the star $10^7$~yr, when the protoplanetary disk completely disappeared. The simulations lasted until the final orbit $a_\mathrm{ref}$ was reached by the planet, but not later than at $5\times10^9$~yr.

Under the action of viscosity, the gas diffuses outward and inward. On the inner side, the gas receives a gravitational torque from the planet, thus forming a gap. In the outer part, a power-law density profile is formed. Photoevapration tends to make the density profile steeper, as can be seen at early times (Fig.~\ref{fig:grid-sigma-1e8}), when the gas density in the torus is still low. It is interesting that an increase of the viscosity parameter $\alpha$ by an order of magnitude leads to a decrease of the surface density in the maximum point, but by no more than three times. Apparently, this is caused by an increase in the diffusion flux to the outer part of the torus. The sensitivity of the density distribution to the viscosity index $\beta$ is less pronounced. Later, the density profile stabilizes and takes the form $\Sigma \propto r^{-(\beta+1/2)}$, see Fig.~\ref{fig:sigma-fin}.

As it turned out, introduction of photoevaporation into the model dramatically affects migration rates (Fig.~\ref{fig:grid-a}). The loss of the planet's ejecta greatly reduces the rate of accumulation of matter in the torus, compared to the model without this effect (Paper~I). In the present model, accumulation competes with the photoevaporation, slowing down the migration. Total migration time may increase by one and a half orders of magnitude in the strong turbulence case ($\alpha = 0.01$, $\beta = 1.5$) and by more than two orders of magnitude in the weak turbulence one ($\alpha = 0.001$, $\beta = 1$).

There are three factors in the proposed model that are responsible for removal of the angular momentum of the system: advection, viscosity, and photoevaporation. The torque to which the torus is subject to can be obtained by integration of Eq.~\eqref{eq:angular-momentum-transfer-general} by the torus' surface:
\begin{multline}
  \label{eq:total-angular-momentum-conservation}
  \dot{J}_\mathrm{t}
  \equiv 2\pi \int_{r_0}^{r_\mathrm{out}} dr\,r\,\pdiff{}{t} (\Sigma r^2 \Omega) =  \\
  = 2\pi\,(r F r^2 \Omega - r^2 W) \bigr|_{r=r_0} - 2\pi\,(r F r^2 \Omega - r^2 W) \bigr|_{r=r_\mathrm{out}}  \\
  - 2\pi \int_{r_0}^{r_\mathrm{out}} dr\,r\,\dot{\Sigma}_\mathrm{pe}\,r^2 \Omega
  + 2\pi \int_{r_0}^{r_\mathrm{out}} dr\,r\,\Sigma \tau  \;.
\end{multline}
Here $r_\mathrm{out} = r_0 x_N$ is an outer edge of the torus in the numerical model. The first two terms at the r.h.s. of Eq.~\eqref{eq:total-angular-momentum-conservation} is the amount of angular momentum passing through the boundaries of the torus per unit time, $\dot{J}_\mathrm{b}$, they include advective and viscous transfer. The third term is the angular momentum spent to the photoevaporation, $\dot{J}_\mathrm{pe}$. The last term is the planet's torque (with the negative sign, $-\dot{J}_\mathrm{p}$). Strictly speaking, the planet's torque must contain also a term $- (r F r^2 \Omega) \bigr|_{r=r_0}$ since this is the rate the angular momentum of the gas leaves the planet's atmosphere. However, this is a small quantity and we will neglect it. In total the eq.~\eqref{eq:total-angular-momentum-conservation} can be reformulated as an explicit conservation law:
\begin{equation}
  \dot{J}_\mathrm{p} + \dot{J}_\mathrm{t} - \dot{J}_\mathrm{pe} - \dot{J}_\mathrm{b} = 0  \;.
\end{equation}
This conservation law is demonstrated in a Fig.~\ref{fig:conservation}. As one can see, at early times the photoevaporation is a main sink of the angular momentum. After $10^8$~yr the boundary flux becomes more important. To the final time nearly all the planet's orbital momentum passes through the outer boundary ($-J_\mathrm{b} > 0$). Only a small part of it is accumulated in the torus.

The similar conservation law can be formulated for the mass:
\begin{multline}
  \label{eq:total-mass-conservation}
  \dot{M}_\mathrm{t}
  \equiv 2\pi \int_{r_0}^{r_\mathrm{out}} dr\,r\,\pdiff{\Sigma}{t} =  \\
  = 2\pi r_0 F_0 - 2\pi r F \bigr|_{r=r_\mathrm{out}}
  - 2\pi \int_{r_0}^{r_\mathrm{out}} dr\,r\,\dot{\Sigma}_\mathrm{pe}  \;.
\end{multline}
The first term at the r.h.s. is the planet's mass sink, $-\dot{M}_\mathrm{p}$. The second term is the mass ouflow through the outer boundary, $\dot{M}_\mathrm{b}$. The last term is the total photoevaporation rate in the torus, $\dot{M}_\mathrm{pe}$. The explicit conservation law is $\dot{M}_\mathrm{p} + \dot{M}_\mathrm{t} - \dot{M}_\mathrm{pe} - \dot{M}_\mathrm{b} = 0$.

We also tested the sensitivity of the dynamics of migration to the variations of the atmosphere outflow rate $\dot{M}_\mathrm{ref}$ and X-ray stellar luminosity $L_\mathrm{X,ref}$. The change of the outflow rate by a factor $2$ led to the change of the migration time (with opposite sign) by about two times (Fig.~\ref{fig:a-var}). The change of the X-ray luminosity by the same factor led to a much weaker change in the migration time (of the same sign). Note, both the outflow rate and X-ray luminosity were varied independently in both tests presented in Fig.~\ref{fig:a-var}. In fact, an increase of luminosity should lead to the proportional increase in the atmosphere outflow rate, as mentioned in Sec.~\ref{sec:outflow}, since $\dot{M}_\mathrm{ref} \propto L_\mathrm{XUV}$. Therefore, the net effect of an increase of X-ray luminosity should be accleration of migration.

The dynamics of the torus mass accumulation depends on the viscosity law. In a short time after the start of the simulations, the torus gains a mass of $10^{23} \mathdash 10^{24}$~g, after which the mass slowly changes by $1 \mathdash 2$ orders of magnitude (Fig.~\ref{fig:grid-mass}). Later, the mass of the torus is limited by the initial orbit of the planet, without significant dependence on viscosity. Migration times, however, differ much. The reason for this is the width of the gap. It is different for different values of the viscosity parameter, as can be  seen already in Figs.~\ref{fig:grid-sigma-1e8} and~\ref{fig:sigma-fin}. In Paper~I we've found that the gap is forming rather quickly, in the viscous timescale. Hence, the width of the gap can be estimated from the balance of the gravitational torque and the viscous torque. For this, we write down the r.h.s. of Eq.~\eqref{eq:angular-momentum-conservation}, assuming that the torus matter is concentrated in close vicinity of the orbit $a + H_\mathrm{gap}$, and $H_\mathrm{gap} \ll a$ (i.e., we apply impulse approximation, see \citet{Papaloizou2006RPPh...69..119P}):
\begin{equation}
  \label{eq:grav-torque-impulse}
  \dot{J}_\mathrm{gr}
  \sim 24C_0 \left( \frac{M_\mathrm{p}}{M_\mathrm{s}} \right)^2 \left( \frac{a}{H_\mathrm{gap}} \right)^3
    a^4 \Omega_\mathrm{p}^2 \Sigma  \;.
\end{equation}
The viscous torque is
\begin{equation}
  \label{eq:visc-torque}
  \dot{J}_\mathrm{visc}
  \sim \nu a^2 \Omega_\mathrm{p} \Sigma
  \sim \alpha H^2 \Omega_\mathrm{p}^2 \Sigma  \;.
\end{equation}
Equality of both expressions gives
\begin{equation}
  \label{eq:gap-width}
  \frac{H_\mathrm{gap}}{a}
  \sim \left[ \frac{24C_0}{\alpha} \left( \frac{M_\mathrm{p}}{M_\mathrm{s}} \right)^2 \left( \frac{a}{H} \right)^2 \right]^{1/3}  \;.
\end{equation}
According to the estimate, the models with $\alpha = 0.001$ have about two times wider gaps, than the models with $\alpha = 0.01$. It would seem that this dependence is insignificant, but it has the strongest effect on the planet migration rate, since the  tidal torque depends on the cube of the gap width, see Eq.~\eqref{eq:grav-torque-impulse} and \citet{Papaloizou2006RPPh...69..119P}. The physical reason for the narrowing of the gap is that, in accordance with  accretion theory \citep[e.g.,][]{Shakura1973A&A....24..337S}, viscosity causes the mass to flow inward.

\section{Discussion and conclusions}

As it is seen  in the Fig.~\ref{fig:grid-a}, the planet spends most of the total migration time in its initial orbit. During the first few  $10^7$~yr, the mass of the torus decreases, only later it begins to accumulate mass stably. Consequently, the success of the migration is highly dependent on the activity of the star during this time, since a sufficiently strong burst of ionizing radiation may evaporate the torus.

\begin{figure}
  \begin{center}
    \includegraphics{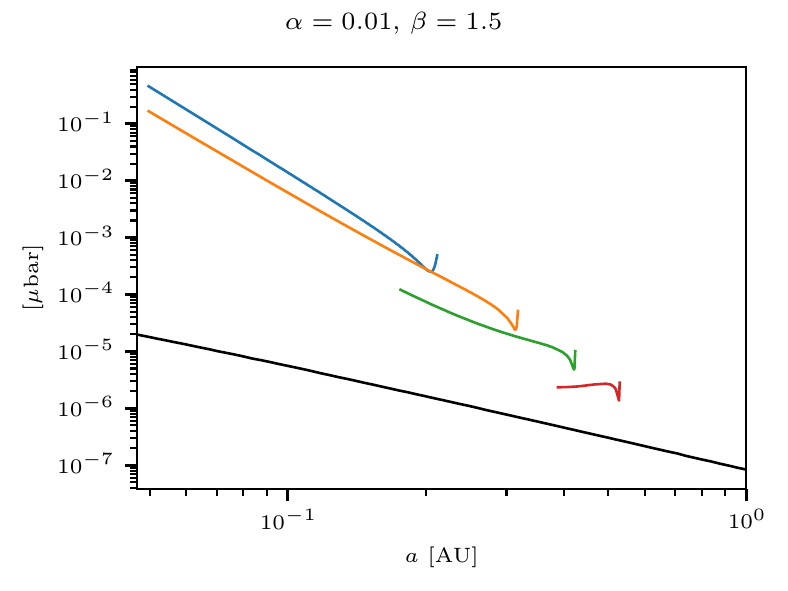}
  \end{center}
  \caption{Checking the condition \eqref{eq:wind-dominance-condition}. {\em Coloured lines}: Pressure (in microbars) of the gas at the inner edge of the torus. The colors of the lines have the same meaning as in Fig.~\ref{fig:grid-a}. {\em Black line}: Dynamic pressure of the solar wind for coronal temperature $2\times10^6$~K according to the \citet{Parker1958ApJ...128..664P} model.}
  \label{fig:wind}
\end{figure}

In Sec.~\ref{sec:stellar-wind} we made rough estimates of the role of the stellar wind and showed that in can be neglected. Based on the results of simulations in this study, we can confirm this conclusion, albeit, with some reservations. Figure~\ref{fig:wind} shows the gas pressure at the inner edge of the torus depending on the radus of the planet's orbit. As it is seen, in all models, gas pressure exceeds the dynamic one of the stellar wind (according to the Parker model). The general trend $P_\mathrm{gas} \propto a^{-5}$ is clearly visible, although in each model separately, the dependence is more complex. According to this trend, for higher orbits, gas pressure in the torus will rapidly decrease, approaching the dynamic pressure of the wind. This means that starting from an orbit of about $1$~AU and for more distant ones, the torus is unlikely to survive under the influence of the stellar wind.

Photoevaporation significantly affects the rate of accumulation of the matter in the torus. An interesting question is how the wind interacts with the planet's atmosphere ejecta at early times, when the density of the matter in the torus is still low. Unfortinately, this problem cannot be solved  within the presented model (but see \citet{Matsakos2015A&A...578A...6M} and \citet{Khodachenko2019ApJ...885...67K}).

Another interesting question is the gas-dynamic interaction between the planet's envelope and the torus. In the protoplanetary disk, the gas of the outer layers of the envelope can be replenished by the material of the disk \citep{Ormel2015MNRAS.447.3512O}. As a result, the heat balance and, generally speaking, the ionization balance change. Potentially, this can lead to a change in the dependence of the rate of outflow on the atmosphere on the ionizing radiation flux.

Of course, we did not take into account all the possible effects. In addition to the stellar wind mentioned above, gravitational interaction with other planets and direct tidal interaction between the star and the planet potentially may have an impact. When the planet's orbit is close to the star, their tidal interaction  can change the eccentricity and semi-major axis of the planet's orbit. The sign of this effect depends on the ratio of the stellar rotation period ($P_\mathrm{rot}$) and the orbital period of the planet ($P_\mathrm{orb}$), and its magnitude strongly depends on the semi-major axis of the orbit \citep[see][and references therein]{Jackson2008ApJ...678.1396J}. For the system HD~209458 characteristic time of the tidal migration assuming circular orbit can be estimated as \citep{Jackson2008ApJ...678.1396J}
\begin{equation}
  \label{eq:direct-tidal-migration-time}
  \left| \frac{a}{\dot{a}} \right|
  \sim (2.6\times10^9 \text{~yr}) \left( \frac{a}{0.04 \text{~AU}} \right)^{13/2}  \;.
\end{equation}
If we assume $a = 0.047$~AU, then $|a/\dot{a}| = 7.3\times 10^9$~yr, i.e., the direct tidal effect may act only at late time, when the planet approaches the star very close due to some other mechanism. In this case, $P_\mathrm{rot} > P_\mathrm{orb}$ (applying, e.g., the empirical rotation-age law of \citet{Barnes2003ApJ...586..464B} to the host star of the system under study), so the planet migrates inward. However, since the tidal effect is very weak, we will ignore it. The account of possible non-zero eccentricity of the planet's orbit ($e \leq 0.02$, \citet{Lanza2010A&A...512A..77L}) only halves the estimate of the migration time \eqref{eq:direct-tidal-migration-time}, which is not interesting.

Based on the simulation results, the following can be stated. For migration of the planet HD~209458b to its current observable orbit at $0.047$~AU during the lifetime of the star, it is necessary for the initial radius to be less than $\sim 0.35$~AU for the `strong' turbulence ($\alpha = 0.01$, $\beta = 1.5$) or $\sim 0.3$~AU for the `moderate' turbulence ($\alpha = 0.001$, $\beta = 1.5$ or $\alpha = 0.01$, $\beta = 1$). So we may conclude that the atmospheric outflow can be important factor for the migration of the hot jupiters.

\section{Acknowledgements}

This study was supported by the Ministry of Science and Higher Education of the Russian Federation under the grant No. 075-15-2020-780 (N13.1902.21.0039). Section 4 was prepared with the support of the Russian Science Foundation (Project No. 18-12-00447).

\section{Data Availability}

Software code and data are available at \textsc{Github} repository via \url{https://github.com/evgenykurbatov/kb21-hotjup-migration-adv}




\bsp  
\label{lastpage}
\end{document}